\documentclass[aps,prl,twocolumn,showpacs,superscriptaddress,groupedaddress]{revtex4}  
\usepackage{graphicx}  
\usepackage{dcolumn}   
\usepackage{bm}        
\usepackage{array}
\usepackage{placeins}
\usepackage{longtable}
\usepackage{placeins}

\usepackage{color}
\usepackage{amsmath, amssymb, graphics, setspace}
\usepackage[bookmarks=false]{hyperref}
\newcommand{\mathsym}[1]{{}}
\newcommand{\unicode}[1]{{}}

\newcommand{\be}{\begin{equation}}
\newcommand{\ee}{\end{equation}}

\newcommand{\rmmax}{{\rm{max}}}
\newcommand{\trap}{{\rm{trap}}}

\newcommand{\bfr}{{\bf{r}}}
\newcommand{\bfv}{{\bf{v}}}

\newcommand{\tbfr}{{\textbf{r}}}
\newcommand{\tbfv}{{\textbf{v}}}

\newcommand{\calO}{{\cal{O}}}

\newcommand{\calG}{{\cal{G}}}
\newcommand{\elliptic}{{\rm{elliptic}}}
\newcommand{\scissor}{{\rm{scissor}}}
\newcommand{\kinetic}{{\rm{kinetic}}}

\newcommand{\LDA}{{\rm{LDA}}}

\newcommand {\E}{\nabla\phi}
\newcommand {\bea}{\begin{eqnarray}}
\newcommand {\eea}{\end{eqnarray}}

\def\beq{\begin{eqnarray}}\def\eeq{\end{eqnarray}}
\def\be{\begin{equation}}\def\ee{\end{equation}}

\begin{document}



\title{A Proposal for measuring Anisotropic Shear Viscosity in Unitary Fermi
Gases}

\date{\today}
\author{Rickmoy Samanta}
\email{rickmoysamanta@gmail.com}
 \affiliation{Department of Physics, Bar Ilan University, Ramat Gan, 52900,
 Israel}
\affiliation{
 Department of Theoretical Physics, Tata Institute of Fundamental Research,
 Dr. Homi Bhabha Road, Mumbai, 400005}
 \email{rickmoysamanta@gmail.com}
 \author{Rishi Sharma}
 \email{rishi@theory.tifr.res.in}
\affiliation{
 Department of Theoretical Physics, Tata Institute of Fundamental Research,
 Dr. Homi Bhabha Road, Mumbai, 400005}
\author{Sandip P. Trivedi}
 \email{trivedi.sp@gmail.com}
\affiliation{
 Department of Theoretical Physics, Tata Institute of Fundamental Research,
 Dr. Homi Bhabha Road, Mumbai, 400005}

\begin{abstract}
We present a proposal to measure anisotropic shear viscosity in a  strongly
interacting, ultra-cold, unitary Fermi gas confined in a harmonic trap. We
introduce anisotropy in this setup by strongly confining the gas in one of the
directions with relatively weak confinement in the remaining directions. This
system has a close resemblance to anisotropic strongly coupled field theories
studied recently in the context of gauge-gravity duality. Computations in such
theories (which have gravity duals) revealed that some of the viscosity
components of the anisotropic shear viscosity tensor can be made much  smaller
than the entropy density, thus parametrically violating the bound proposed by
Kovtun, Son and Starinets (KSS): $\frac {\eta} {s} \geq \frac{1}{4 \pi}$. A
Boltzmann analysis performed in a system of weakly interacting
particles in a linear potential also shows that components of the viscosity
tensor can be reduced. Motivated
by these exciting results, we propose two hydrodynamic modes in the unitary
Fermi gas whose damping is governed by the component of shear viscosity
expected to violate the KSS bound. One of these modes is the well known scissor mode. We
estimate trap parameters for which the reduction in the shear viscosity is significant and find that the trap geometry, the damping timescales, and mode
amplitudes are within the range of existing experimental setups on ultra-cold
Fermi gases.
\end{abstract}

\maketitle

\textit{Introduction -} The computation of transport properties of strongly
interacting quantum field theories is a challenging problem and has attracted
physicists working on a wide variety of systems including ultra-cold Fermi gases
at unitarity~\cite{Adams:2012th,Bulgac:2010dg,Taylor:2010}, heavy ion
collisions~\cite{Adams:2012th,Bhalerao:2010wf,CasalderreySolana:2011us}, and neutron
stars~\cite{Page:2006ud,Alford:2007xm}.  


The AdS/CFT correspondence~\cite{Maldacena:1999} has provided many insights
into the transport properties of such strongly coupled field theories which
have gravity duals.  In the limit of large t'Hooft coupling $\lambda$ and large
number of ``colors'' $N_c$, the dual gravity theories are effectively classical
and the computation of transport properties become much easier. One finds that
in all such isotropic strongly coupled theories in $3+1$ dimensions which admit
smooth gravity duals, the ratio of shear viscosity $\eta$ to entropy density $s$ is
${\eta \over s} = \frac{1}{4\pi}$~\cite{Son:2002sd,Kovtun:2004de} (we work in
units where $\hbar=c=1$). Weakly coupled theories typically have much larger
${\eta \over s}$. This led Kovtun, Son and Starinets (KSS) to conjecture that
${\eta \over s}$ is bounded from below by $1/(4\pi)$. It was later found that
finite $\lambda$ corrections (which correspond to higher derivative corrections
in the gravity side) can drive ${\eta \over s}$ below the KSS
bound~\cite{Brigante:2007nu,Brigante:2008gz,Kats:2007mq,Buchel:2008vz,Sinha:2009ev,Cremonini:2011iq}.

The gravity duals of ultra-cold Fermi gases and quark gluon plasma produced in
heavy ion collisions are not known. However, beautiful experiments have managed
to measure the value of $\eta/s$ in these two systems. In each of these cases,
the measured value of $\eta/s$ is close to $1/(4\pi)$. For example, the value
of $\eta/s$ has been measured for ultra-cold fermions at unitarity for a wide
range of temperatures and the minimum value (see
Refs.~\cite{Schafer:2007pr,Cao58,Thomas:2015}) is about six times the KSS
bound.    

The above results are for isotropic phases. But several phases in nature are
anisotropic, for example spin density waves, spatially modulated phases and
phases where anisotropy arises due to an external field. It is thus interesting
to study the behavior of the different components of the resultant anisotropic
shear viscosity tensor in such strongly coupled anisotropic systems. The
situation we explore here features an externally applied field in a particular
direction which gives rise to anisotropies in the shear viscosity. This
possibility is well studied in weakly coupled theories in the presence of a
background magnetic
field~\cite{Landau1987Fluid,Tuchin:2011jw,Ofengeim:2015qxz}. However the study
of transport coefficients in strongly coupled field theories is less explored.
Recently, anisotropic gravitational backgrounds in field theory have been
studied using the AdS/CFT correspondence (see
\cite{Landsteiner:2007bd,Azeyanagi:2009pr,Natsuume:2010ky,Erdmenger:2010xm,
Basu:2011tt,Erdmenger:2011tj,Mateos:2011ix,Mateos:2011tv,Iizuka:2012wt}) and the computation
of the  viscosity in  some  of these anisotropic phases has also been performed
(see \cite{Rebhan:2011vd, Polchinski:2012nh} and
\cite{Giataganas:2012zy,Mamo:2012sy,Jain:2014vka,
Critelli:2014kra,Ge:2014aza,Jain:2015txa}). 

The results of Ref.~\cite{Jain:2014vka} and
Ref.~\cite{Jain:2015txa} indicate that parametric violations of the KSS bound
are possible in such anisotropic scenarios. It was shown in
Ref.~\cite{Jain:2015txa} that this behavior is quite general and arises in
situations where the force responsible for breaking of isotropy is spatially
constant. By increasing the anisotropy compared to the temperature, the ratio
for appropriate components of the shear viscosity to entropy density can be
made arbitrarily small, violating the KSS bound. \\
It is natural to inquire if this behaviour can be observed in experiments on
strongly coupled systems under appropriate conditions. While the systems
studied in Refs.~\cite{Jain:2014vka,Jain:2015txa} can not be created in the
laboratory, the ability to tune the thermodynamics and geometry of trapped
ultra-cold fermions at unitarity~\cite{Cao58,
PhysRevLett.108.070404,Thomas:2015, PhysRevLett.99.150403, 2016arXiv160703221R}
makes them ideal candidates to see such effects.

In this paper we demonstrate that in the presence of an anisotropic trap and an
appropriate choice of parameters one can create ultra-cold Fermi systems which
share many essential features of the theories considered in
Refs.~\cite{Jain:2014vka,Jain:2015txa}.  We give a concrete proposal for the
trap geometry and parameters where a parametric violation of the KSS bound is
likely to be seen.

\textit{ Gravity Results -} We briefly review results of computations 
of shear viscosity in the gravity picture obtained by studying anisotropic
blackbranes~\cite{Jain:2014vka} where the breaking of isotropy is due to an externally applied
force which is translationally invariant. The simplest system
discussed in Ref.~\cite{Jain:2014vka} consists of a massless dilaton
minimally coupled to gravity, and a cosmological constant. The action is 
\begin{equation}
S = \frac{1}{16\pi G} \int d^5 x \sqrt{g}~ [R+12\Lambda -
\frac{1}{2}\partial_\mu \phi\partial^\mu\phi]\;,~\label{eq:5dlag}
\end{equation}
where $G$ is Newton's constant in $5$ dimensions and $\Lambda$ is a cosmological
constant. The dual field theory in the absence of anisotropy is a $3+1$ dimensional
conformal field theory. The dilaton profile, linear in the spatial co-ordinate
$z$
\begin{equation}
\label{anisoparam}
\phi=\rho z\;,
\end{equation}
explicitly breaks the symmetry to $2+1$.\\
Using AdS/CFT one finds~\cite{Jain:2014vka} that for a system at temperature
$T$, (using the compact notation $\eta_{ijij}=\eta_{ij}$) $\eta_{xz}=\eta_{yz}$
(which are spin 1 with respect to the surviving Lorentz symmetry) is affected by
the background dilaton. In the low anisotropy regime ($\rho/T \ll 1$):
\begin{equation}
\frac{\eta_{xz}}{s}=
\frac{1}{4\pi}-\frac{\rho^2 \log 2}{16 \pi^3 T^2}
+
\frac{(6-\pi^2+54 (\log 2)^2)\rho^4}{2304\pi^5 T^4}
+
{\cal{O}}\bigg[\bigg(\frac{\rho}{T}\bigg)^6\bigg]~\label{eq:eta_low}\;.
\end{equation}
The correction to the zero anisotropy result, the KSS bound
$\frac{\eta_{xz}}{s}=\frac{1}{4\pi}$, is proportional to $\frac{(\nabla
\phi)^2}{T^2}$ where $\nabla \phi=\rho\hat{z}$ is the driving force and $1/T$
is the microscopic length scale in the system. \\
In extreme anisotropy ($\rho/T \gg 1$),
\begin{equation}
{\eta_{xz}}/{s}\rightarrow ({1}/{4\pi}) ({32\pi^2 T^2}/{3\rho^2})\,
\end{equation}
and hence becomes parametrically small~\cite{Jain:2014vka}. But this domain
will not be physically accessible in the cold atom systems.\\
In contrast the $\eta_{xy}$ component (which couples to a spin $2$ metric
perturbation) was found to be unchanged from its value in the isotropic case, 
$\frac{\eta_{xy}} {s}= \frac{1}{4\pi}$.\\
Parametric reduction of the spin $1$ components of $\eta/s$ has been found for
a variety of strongly coupled theories with a gravitational
dual~\cite{Jain:2015txa,Rebhan:2011vd}. \\
Motivated by the above results, we may expect to observe parametrically
suppressed viscosities compared to the KSS bound in systems with the following
properties:
\label{cond}
\begin{enumerate}
\item{It is strongly interacting and in the absence of anisotropy has
a viscosity close to the KSS bound.}
\item{The equations of hydrodynamics admit modes whose damping is sensitive to
the spin 1 viscosity components
Ref.~\cite{Jain:2014vka,Jain:2015txa}.}
\item{The gradient of the background potential (say in the
$z$ direction) must be significant compared to a microscopic scale governing
transport}
\item{The background potential responsible for breaking of isotropy is approximately
spatially constant.}
\item{The velocity gradients are small enough to ensure validity of hydrodynamics.}
\end{enumerate}

\textit{ Ultra-cold atoms-} Ultra-cold unitary Fermi
gases~\cite{Stringari:2008,Ketterle:2008} are strongly interacting systems with
one of the lowest $\eta/s$~\cite{Schafer:2007pr,Cao58,Thomas:2015} measured,
and (based on the above criteria) a good candidate system to explore small anisotropic
viscosities. 

Typically they are trapped in harmonic potentials 
\begin{equation}
\phi(\bfr) =  \sum_i   m \omega_{i}^{2} x_{i}^{2}/2
~\label{eq:harmonic_potential}
\end{equation}
where $i$ runs over the $x,y,z$ directions and $m$ denotes the mass of the
fermionic species which we take as $^6$Li. The trap frequencies are chosen such that
$\omega_z\gg\omega_x, \omega_y$ the potential gradient is dominantly in the $z$
direction. In Section \textit{Anisotropy} we show how large $\omega_z$ is
required to measure significant deviations from the isotropic shear viscosity.

We note that while ultra-cold Fermi gases share some important features with
the theories considered in~\cite{Jain:2014vka,Jain:2015txa}, there is an
important difference, namely, unlike the field theories
of~\cite{Jain:2014vka,Jain:2015txa} the stress energy tensor in a trap is not
translationally invariant. Even so, the Boltzmann equation (see
Eqs.~\ref{eq:anisotropy_corrections0}, \ref{eq:etaoflambda2}) for ultra-cold
Fermi gases also predicts a reduction of $\eta_{xz}$.\\
%
\textit{ Hydrodynamic modes-} 
We find two solutions of equations of superfluid hydrodynamics in a harmonic
trap, which are sensitive to the spin 1 components of the viscosity tensor.
Each of these modes is characterized by the superfluid and the normal
components, which we denote by $\textbf{v}_s$ and $\textbf{v}_n$ respectively.

The first mode, which we call the elliptic mode, has $\textbf{v}_s=0$ and
$\textbf{v}_n=\textbf{v}$ given by
\begin{equation}
\label{eq:vprofile}
{\bf{v}} =e^{i \omega t} (\alpha_x z ~\hat{x} + \alpha_z
x \hat{z})
\end{equation}
with the following relations:
\begin{equation}
\label{modea}
\elliptic:\;\;\omega=0,~
   \alpha_{z}=-\alpha_{x} {\omega_{x}^{2}}/{\omega_{z}^{2}} 
\end{equation}

The second mode, the well known scissor mode, has 
$\textbf{v}_s=\textbf{v}_n=\textbf{v}$ given by Eq.~\ref{eq:vprofile} with
\begin{equation}
\label{scissor}
\scissor:\;\;
\omega=\sqrt{\omega_x^{2}+\omega_z^{2}},~
   \alpha_{z}= \alpha_{x}.
\end{equation}

We see that in the high anisotropy limit $\omega_{z}\gg\omega_{x}$,
$\alpha_{z}\to 0$ for the elliptic mode, and hence we recover a flow profile
similar to that considered in~\cite{Jain:2014vka}. The elliptic mode has not
been studied in ultra-cold gas experiments. The scissor
mode~\cite{PhysRevLett.83.4452} has been studied extensively in bosonic (for
example see Refs.~\cite{PhysRevLett.84.2056}) fermionic
gases~\cite{PhysRevLett.99.150403}; therefore we focus on it. For both modes,
$\partial_j\tbfv_j=0$. 

To ensure that hydrodynamics is valid in the region where the energy
loss due to viscous damping is substantial, we impose the condition 
\begin{equation}
\alpha_{x}<\alpha_{x}^{\rm{max}} =  
{P ({z_{\rmmax}})}/{\eta_{xz}({z_{\rmmax}})}
~\label{eq:hydro_condition}\;,
\end{equation}
where $z_{\rmmax}$ is the place where the local chemical potential equals the
temperature (Eq.~\ref{eq:zmax}). This sets the upper limit on the amplitude of the
modes. 

\textit { Viscous damping -} The energy dissipated due to the shear viscosity 
is given by
\begin{equation}
\begin{split}
&\dot{E}_{\kinetic} = 
   -\int d^3{\bfr}\, \frac{\eta_{ij}({\bfr})}{2} \, 
  \left(\partial_i\tbfv_j+\partial_j\tbfv_i-\frac{2}{3}\delta_{ij}
      \partial_k \tbfv_k \right)^2\;. 
   ~\label{eq:dissipation}
\end{split}
\end{equation}
$T$ is constant in the elliptic and scissor modes and hence thermal
conduction is neglected.

For the elliptic mode, 
\begin{equation}
\begin{split}
\dot{E}_{\kinetic} &= 
      - \int d^3{\bfr}\, \eta_{xz}({\bfr}) ~\alpha_{x}^{2} 
      (1- {\omega_{x}^{2}}/{\omega_{z}^{2}})^2\;,
      \label{etaexp}
\end{split}
\end{equation}
and for the scissor mode 
\begin{equation}
\begin{split}
\dot{E}_{\kinetic} &= -2\int d^3{\bf{r}}\, \eta_{xz}({\bf{r}}) ~\alpha_{x}^{2} .
 \label{etaosc}
\end{split}
\end{equation}

\begin{figure}
   \begin{center}
  \includegraphics[width=9cm]{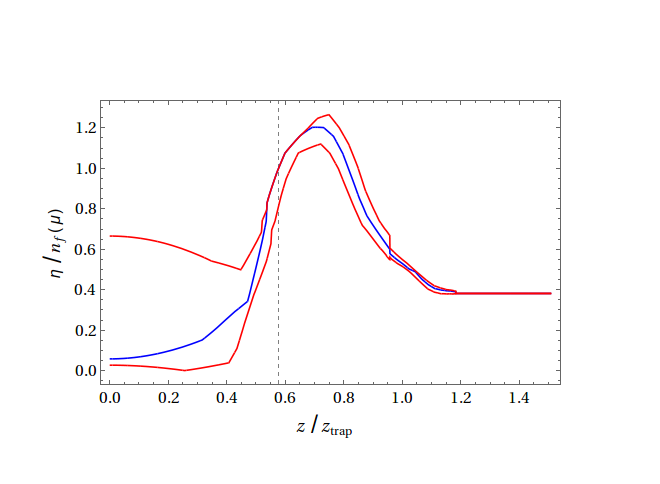}
   \end{center}
    \caption{(Color online) Local shear viscosity in units of $n_f$ at the
    center (Eq.~\ref{eq:n_fmu}) as a function of $z$ in units of $z_{\trap}$
    (Eq.~\ref{eq:ztrap}) for $T={2T_{c}/3}$. The red curves denote the error
    estimate which include errors in the measurement of
    $\eta/n$~\cite{Thomas:2015} as well as errors in $\calG$ due to errors in
    the measurements of thermodynamics~\cite{Ku563}.  The black dashed vertical
    line is at $z_c$. } \label{localvis}
\end{figure}

The total mechanical energy $E$ is twice the average
kinetic energy ($E=2{E}_{\kinetic}$) where,
\begin{equation}
\begin{split}
{E}_{\kinetic} &= 
      \langle \frac{1}{2} \int d^3{\bfr}\, m n({\bfr}) \bfv^2\rangle\;,
      \label{energy}
\end{split}
\end{equation}
where $\bfv$ is the velocity of either mode and the average is taken over one
cycle for the scissor mode (the elliptic mode is non-oscillatory).  

In the strong anisotropy limit $ \omega_z \gg \omega_x$, it can be shown that
the energy of the elliptic mode scales as $E_{\elliptic}\sim \frac{\mu^6}{\omega_x \omega_y
\omega_z^3} $ and that of the scissor mode scales as $E_{\scissor}\sim \frac{\mu^6}
{\omega_x^3 \omega_y \omega_z}$. (The scalings of the scissor mode, for
example, can be derived as follows: 
$ E \sim \int dx dy dz [m n v^2] \sim  L_x L_y L_z [m n \alpha^2 L_x^2] \sim \frac{\mu^6} {\omega_x^3
\omega_y \omega_z} $, where we have assumed that at the center of the trap
$\mu>0$ and $L_i=\sqrt{2\mu/(m\omega_i^2)}$.) In a similar manner, one can derive the approximate
scalings for energy dissipation rates:  $\dot{E} \sim \frac{\mu^5}{\omega_x
\omega_y \omega_z}$ for both the modes (assuming $\eta$ scales the same way as
$n$ ie. $\sim ( m \mu)^{3 \over 2}$).
    
\textit{Thermodynamics and trap results -} The evaluation of the energy loss
from Eq.~\ref{etaexp} and Eq.~\ref{etaosc} requires the viscosity $\eta$ as a
function of the position ${\bfr}$ in the trap. To estimate $\eta(\bfr)$ we use
the local density approximation (LDA).\\

For the unitary Fermi gas in the thermodynamic limit, the
thermodynamic functions can be factorized into universal dimensionless
functions~\cite{Schafer:2007pr} and an overall scale given, for example, in terms
of the number density of a free Fermi gas,
\begin{equation}
n_f(\mu) = (2m\mu)^{3/2}/{(3\pi^2)}~\label{eq:n_fmu}\;.
\end{equation}

In particular,
\begin{equation}
\begin{split}
n(\mu,\;T)=&n_{f}(\mu)[\mathcal{G}(y)-2y\mathcal{G}^{\prime}(y)/5], \\
s(\mu,\;T)=&({2}/{5})n_{f}(\mu)\mathcal{G}^{\prime}(y)~\label{nands},
\end{split}
\end{equation}
where 
\begin{equation}
y={T}/{\mu}~\label{eq:y}\;.
\end{equation}
$\mathcal{G}(y)$ can be obtained (see Ref.~\cite{Samanta:2016pic} for details)
by analyzing the thermodynamic measurements carried out in Ref.~\cite{Ku563}.
Similarly, the ratios $\eta/n$ and $\eta/s$ are dimensionless functions of
$y$~\cite{Schafer:2007pr,Cao58,Thomas:2015}. 

The local value of the chemical potential in a harmonic trap
(Eq.~\ref{eq:harmonic_potential}) is 
\begin{equation}
\mu({\textbf r}) = \mu - \phi({\textbf r})\;,
\end{equation}
($\mu$ without the argument ${\textbf{r}}$ is the chemical potential at the
center of the trap). $T$ is independent of $\tbfr$.

At temperatures much smaller than the chemical potential, transport is
dominated by the Goldstone mode associated with superfluidity and the viscosity
can be computed in a controlled manner by using an effective
theory~\cite{PhysRevA.76.053607}. At temperatures large compared to the
chemical potential or when the chemical potential is negative, the density of fermions is small and a kinetic calculation
of the viscosity, $\eta={\rm{const.}}\times(mT)^{3/2}$, is
reliable~\cite{Bruun:2005en,Bruun:2006kj,Bluhm:2015raa,Enss:2010qh} and
consistent with the experiments~\cite{Thomas:2015}. In the region just above
$T_c\approx0.4\mu$, a theoretical evaluation of the viscosity is difficult.
Monte Carlo~\cite{Wlazlowski:2012jb,Wlazlowski:2015yga} methods, microscopic
approaches~\cite{Guo:2010dc}, and $T-$matrix techniques~\cite{Enss:2010qh} have
been used to calculate the viscosity in this regime but presently the best
estimate for the viscosity in this intermediate regime comes from
experiments~\cite{Thomas:2015}.
To estimate $\eta({\textbf r})$ in LDA, we find the number density at
${\textbf r}$ using Eqs.~\ref{nands},~\ref{eq:y} for the local $\mu({\textbf r})$ and $T$
and multiply it by the ratio $\eta \over n$ which was measured for a large
range of values of $y$ in Ref.~\cite{Thomas:2015}. 

The qualitative behavior of the viscosity as a function of the distance from
the center of the trap depends on the value of $T/\mu$. We define
$T_c\approx0.4\mu$ as the critical temperature at the center of the trap. If
$T$ is below $T_c$, the center of the trap is in the superfluid state. Moving 
away from the center (in the $z$ direction for concreteness),
$\mu(z)$ decreases and at some distance $z=z_c$ there is transition from
the superfluid to the unpaired phase.

{\small
\begingroup
\squeezetable
\begin{table}
    \begin{center}
        \begin{tabular}{|c|c|c|c|c|c|c|c|}
            \hline
             $T$          
       
           & $\frac{z_{0}}{z_{\rm{trap}}} $
           & $l$
           & $\frac{T}{\mu(z)}|_{z_0}$
           & $\frac{\eta}{n}|_{z_0}$
           & $\frac{\eta}{s}|_{z_0}$
           & $\kappa_{{\rm{LDA}}}\frac{10 \mu
           {\rm{K}}}{\mu}\frac{\omega_z/(2\pi)}{10^4{\rm{Hz}}}$\\
           \hline 
           $4T_{c}/5$   &0.63 &0.98  &0.54 & 0.89 &0.85 &0.05 \\
           $2T_{c}/3$    &0.71 &0.62   &0.54 & 0.89   &0.85 &0.08 \\
           $4 T_{c}/7$     &0.76 &0.46  &0.54 & 0.89   &0.85 &0.11 \\
           $T_{c}/2$     &0.8 &0.37   &0.55 & 0.91   &0.85 & 0.13\\
           \hline
        \end{tabular}
    \end{center}
    \caption{Trap characteristics for various $T/T_c$. The scaling behavior of $\kappa_{{\rm{LDA}}}$
     with $\omega_z$ is also shown. The entries  were
    calculated for $\mu=10\mu$K, $T_c=0.4\mu$. $l=\frac{\delta z }{z_0}$  tests
    how well the potential can be approximated as a linear potential in the
    regime of interest. $\kappa_{\rm{LDA}}$ (Eq.~\ref{defkappalda}) tests how well
    LDA is expected to work at $z_0$. }
    \label{ptrap1}
\end{table}
\endgroup
}

If $T$ is just below $T_c$ the plot of the local shear viscosity as a function
of $z$ exhibits a peak at $z=z_0$ close to $z_c$. Qualitatively we understand
this from the fact that the local entropy (see Eq.~\ref{nands}) is the product
of $n_f(\mu({\textbf{r}}))$ which decreases as we move away from the center,
while the function $\mathcal{G}^{\prime}(T/\mu({\textbf{r}})))$
increases~\cite{Samanta:2016pic}. Therefore $s$ as a function of $z$ naturally
has a peak. Since the ratio of the shear viscosity to the entropy density is a
relatively slowly varying function of $z$ in the region around the phase
transition~\cite{Schafer:2007pr}, we also expect the local shear viscosity to 
show a peak. 

An illustrative example for $T=2T_c/3$ is shown in Fig.~\ref{localvis}. We
have defined
\begin{equation}
\begin{split}
z_{\trap} = \sqrt{2\mu/(m\omega_z^2)}\;,~\label{eq:ztrap}
\end{split}
\end{equation}
to scale the $z$ axis and use $n_f(\mu)$ to scale the viscosity. Therefore the
figure is independent of $\mu$ and $\omega_z$ if $T$ is scaled with $\mu$ to
maintain $T=2T_c/3$. Similar behavior is seen for $T$ between $T_c/2$ and $T_c$
and the results are summarized in Table~\ref{ptrap1}.  

Viscous damping dominantly (Eq.~\ref{etaosc}) arises from a
region of width $\delta z$ near $z_0$. This width can be made narrow by
lowering the temperature, such that $l=\delta z/z_0<1$ and in this region the
potential can be approximated as linear~\cite{Samanta:2016pic}. We don't
consider $T$ below $T_c/2$ because for these $T$, the superfluid phonon
viscosity~\cite{PhysRevA.76.053607} is important.

\begingroup
\squeezetable
\begin{table*}[!htbp]
   \begin{center}
        \begin{tabular}{|c|c||c|c|c||c|c|c|}
            \hline
             $T$          
           & $\alpha_x^{\rm{max}}$($10^{-10}$eV) 
           & ${\dot{E}_{\kinetic}}$(j/s)(\textbf{a})
           & $E$(j) (\textbf{a}) 
           & $\tau_0(s)$(\textbf{a})
           & ${\dot{E}_{\kinetic}}$(j/s)(\textbf{b})
           & $E$(j) (\textbf{b}) 
           & $\tau_0(s)$(\textbf{b})\\
           \hline 
           $ 4T_{c}/5$    &$2.83$   &$2.37 \times 10^{-16}$  &$3\times 10^{-20}$    &$0.0002$ & $4.7 \times 10^{-16}$ & $10^{-17}$ &0.04 \\
           $2T_{c}/3$     &$2.35$    &$1.25 \times 10^{-16}$ &$2\times10^{-20}$     &0.0003 & $2.5\times 10^{-16}$ &6.8 $\times 10^{-18}$ &0.05 \\
           $4 T_{c}/7$    &$2.02$   &$7.12 \times 10^{-17}$ &$1.4 \times 10^{-20}$  &0.0004 & $1.4\times 10^{-16}$ &4.8 $\times 10^{-18}$ &0.07 \\
           $T_{c}/2$      &$1.77$    &$4.33\times10^{-17}$    & $1.1\times 10^{-20}$   &0.0005 &$8.65\times 10^{-17}$  &3.6 $\times 10^{-18}$ &0.08\\
           \hline
        \end{tabular}
   \end{center}
    \caption{Energy scales for various $T/T_c$ at  $\omega_z
    =2 \pi \times 10^4$ rads/s, $\omega_x=\omega_y= 2\pi \times 385$ rads/s and
    $\mu=10\mu$K. \textbf{a} denotes the elliptic mode and \textbf{b} the
    scissor mode. The mechanical energy ($E$, see Eq.~\ref{energy}) is given in joules (j) and energy loss rate in
    joules per second, (j/s). For a fixed $ T/ \mu$,  the energy of the
    elliptic (scissor) mode scales as $\frac{1}{\omega_x \omega_y \omega_z^3} $
    ($\frac{1} {\omega_x^3 \omega_y \omega_z}$).  The decay time
    $\tau_0=2E/\dot{E}_{\kinetic}$ [in seconds (s)] of the elliptic (scissor) mode
    scales as $\frac{\mu}{ \omega_z^2} $ ($ \frac{\mu} {\omega_x^2 }$).
    }
    \label{ptrap2}
\end{table*}
\endgroup

All columns in Table~\ref{ptrap1} except $\kappa_{\rm{LDA}}$ (which we will
discuss in the next section) are independent of $\mu$ and $\omega$.

The upper bound on the energy is set by the amplitude $\alpha^{\rmmax}$
(Eq.~\ref{eq:hydro_condition}) with $z_\rmmax$
\begin{equation}
z_{\rmmax}= \sqrt{{2(\mu-T)}/{m \omega_z^2}}~\label{eq:zmax}\;.
\end{equation} 

To estimate energy scales for typical experiments, we take
$T, \mu, \omega_z$ close to those used in experiments in Ref.~\cite{Cao58}. The parameter 
values  used in similar experiments in Refs. \cite{PhysRevLett.99.150403, PhysRevA.78.053609}are comparable, 
but typically the values of $\mu$ and $\omega$ are somewhat smaller.
The maximum angular amplitude of the scissor mode is given by 
\begin{equation}
\begin{split}
\theta \approx
\tan^{-1}\left(({e^{\frac{2\alpha_x }{\omega_z} }-1})
/({e^{\frac{2\alpha_x }{\omega_z}}+1})\right)~\label{eq:theta}\;,
\end{split}
\end{equation}
For $\alpha_x^{\rm{max}}\sim 10^{-10}$ eV (Table~\ref{ptrap2}) and
$\omega_z=2\pi\times10^{4}$ rads/s $\equiv4.16\times10^{-11}$ eV, we find
$\theta_{\rm{max}}\approx45^\circ$. This is larger than the angular
amplitudes measured in~\cite{PhysRevLett.99.150403} and hence within
experimental capabilities. 

The (amplitude) damping time $\tau_0$ defined as
\begin{equation}
\tau_0 = {2E/\dot{E}_{kinetic}}
\end{equation}
is independent of $\alpha^{\rmmax}$ and $\sim10^{-2}(\mu/10\mu K)(2\pi\times
385{\rm{Hz}}/\omega_x)^2$s in the strong anisotropy limit. For $\mu=10\mu K$, $\omega_x=\omega_y=2\pi\times
385$ rads$/$s and $\omega_z=2\pi\times 10^4$ rads$/$s, $\tau_0$ ranges from
roughly $0.04$ to $0.08$. The damping of the scissor mode has been observed for
slightly different parameters values, $\mu\approx 1\mu K$, $\omega_x=2\pi\times 830$ Hz, $\omega_y=2 \pi \times 415$
Hz and $\omega_z=2\pi \times 22$ Hz in Ref.~\cite{PhysRevLett.99.150403} where
the damping time scales measured are of the order of milliseconds. A direct comparison using our technique can only be made for the lowest
temperature ($T/T_F=0.1$) of Ref.~\cite{PhysRevLett.99.150403}. Our
calculations (using the trap parameters of \cite{PhysRevLett.99.150403})give a damping rate of 250 $s^{-1}$ which agrees with experiments \cite{PhysRevLett.99.150403}.

\textit{ Anisotropy -} In LDA the shear viscosity tensor is locally isotropic. If the 
length scale associated with the background potential is shortened so that 
it approaches microscopic length scales, the microscopic properties of the fluid can
become anisotropic. We want to emphasize that this does not automatically 
imply violation of hydrodynamics: as long as the contribution of the viscosity
term to the stress tensor (Eq.~\ref{eq:hydro_condition}) is small, we expect
hydrodynamics to be a good approximation.

In particular, here we shall focus on the anisotropy in the shear viscosity
tensor. In the holographic system mentioned in Sec.~\textit{Gravity Results}, the
microscopic length scale is inverse temperature and the anisotropy is governed by the driving
force $\nabla \phi$ leading to fractional corrections to the viscosity of the order
$(\nabla\phi)^2/T^2$. In the unitary Fermi gas in the unpaired phase ($T>T_c$),
there are two relevant microscopic length scales: the inter-particle separation
($n^{-1/3}$) and the mean free path $\lambda$. Intuitively, we expect corrections 
to the thermodynamic quantities to be proportional to $(\nabla \phi)^2n^{-2/3}$
while the correction to transport properties to be proportional to $(\nabla
\phi)^2\lambda^{2}$.  In the region of interest, $T$ between $T_c$ and $1.5T_c$
where the viscosity contribution in the trap is peaked, it is difficult to calculate these corrections
from first principles since the mean free path is comparable to the
inter-particle separation and the Boltzmann equation is not accurate. In the
absence of controlled techniques in this regime, in order to
estimate these corrections, we solve the Boltzmann equation in the presence a
linear background potential in the relaxation time approximation. The relaxation
time $\tau=\lambda/c_s$ where $c_s$ is the speed of sound. $\lambda$ is treated
as a parameter which depends on the Fermi energy $E_F=(3\pi^2 n)^{2/3}/(2m)$
and the temperature $T$. The result for a weakly interacting Fermi gas for arbitrary $T/E_F$,
is given in Appendix B of~\cite{Samanta:2016pic}. Since we are interested in $T\approx 0.16 E_F$, the
thermal integrals simplify and we obtain the compact result
\begin{equation}
\begin{split}
\eta_{xz}=\eta_{yz} &= \eta[1 + 
    c_2 (\lambda k_F)^2\kappa_{\rm{LDA}}^2+\calO((\lambda\E/\mu)^4)]
\label{eq:anisotropy_corrections0}
\end{split}
\end{equation}
where $k_F = (3\pi^2 n)^{1/3}$, 
\begin{equation}
\begin{split}
c_2 = -{11}/{28}\;.
~\label{eq:etaoflambda2}
\end{split}
\end{equation}
and,
\begin{equation}
\label{defkappalda}
\kappa_{\LDA}= {\nabla \phi}/{(\mu k_F)}|_{z_0}\;. 
\end{equation}
While we will focus on the spin $1$ component, we note that the corrections to
different components of $\eta$ are different and the shear viscosity tensor is
indeed anisotropic~\cite{Samanta:2016pic}. 

In the absence of potential one obtains the well known result~\cite{Ofengeim:2015qxz}
\begin{equation}
\eta = {k_F^4\lambda}/{(15\pi^2)}~\label{eq:eta0}\;.
\end{equation}
By matching Eq.~\ref{eq:eta0} with $\eta/n$ at $z_0$ we find $\lambda
k_F\approx1$ (which verifies the intuition that the coupling is strong and the
Boltzmann calculations are not quantitatively reliable). The corrections are
governed by $\kappa_{\LDA}$.

In the strongly coupled regime for the unitary Fermi gas, $c_2$
(Eq.~\ref{eq:anisotropy_corrections0}) cannot be computed reliably. But it is
intriguing that the weak coupling Boltzmann analysis
(Eq.~\ref{eq:etaoflambda2}) gives $c_2<0$ like the strongly coupled theories
with gravity duals (Eq.~\ref{eq:eta_low},see also Ref.~\cite{Jain:2014vka,Chakraborty:2017msh}).

So far most experiments on hydrodynamics of trapped Fermi gases have been done
for $\kappa_{\LDA}\ll1$~\cite{Cao58,PhysRevLett.99.150403}. For fixed $T/T_c$, 
$\kappa_{\LDA}$ scales as $\omega_z/\mu$ and anisotropic viscosities can be
explored by using larger $\omega_z$ or smaller $\mu$ (or both). For example,
for $T= \frac{T_c}{2}$ (Table~\ref{ptrap1}) and $\mu=10\mu$K (central density
roughly $10^{14}$ atoms/cm$^3$), by increasing $\omega_z$ from typical values
of $2\pi\times10^{4}$ to $8$ times this value, $\kappa_{\LDA}$ can be
increased from $\sim 0.13$ to $\sim 1$. We expect this to lead to an order
unity reduction in $\eta_{xz}$ (Eq.~\ref{eq:etaoflambda2}). 

The damping time for the scissor mode scales as $\mu/(\omega_x)^2$ and
can be kept in the experimentally accessible range of about a millisecond (see
Table~\ref{ptrap2}) while increasing $\omega_z$ (by keeping $\mu$ and 
$\omega_x$ same). This reduces the maximum amplitude $\theta_{\rm{max}}$
(Eq.~\ref{eq:theta}) to $28^\circ$, which is
still in the observable range~\cite{PhysRevLett.99.150403}. 

\textit{Conclusions -} We give the first proposal to measure parametrically
suppressed anisotropic viscosity components in ultra-cold Fermi gases. Our
analyses is motivated by the calculation of the viscosity in strongly coupled
theories with gravitational duals in the presence of a linearly growing
external potential.  The spin $1$ components of the viscosity in these systems
are parametrically reduced from the KSS bound~\cite{Jain:2014vka,Jain:2015txa}.

Our proposal involves a unitary Fermi gas in an anisotropic harmonic trap. We
find that for the temperature at the center of the trap between $0.2$ to $0.4$
times $\mu$, the damping of oscillatory modes is dominated by a region where
the background harmonic potential can be approximated as linear. AdS/CFT then
suggests a reduction in the spin $1$ component of the shear viscosity.

For $\mu=10\mu$K, $T= \frac{T_c}{2}$ ($T_c\approx0.4\mu$), and $\omega_z \sim 2
\pi \times 77000$ rad/s, we find $\kappa_{\LDA} \sim 1$. A Boltzmann analysis
in this regime also predicts an order unity reduction in spin $1$ shear viscosity
components (Eq.~\ref{eq:anisotropy_corrections0}).

Two hydrodynamic modes, an elliptic mode and the well known scissor mode, are
sensitive to this reduction in viscosity. The angular amplitudes
and the decay times are comparable to those measured in~\cite{PhysRevLett.99.150403}.

In the extreme situation for where $\kappa_{\LDA}\sim 1$, our theoretical
estimate for the correction to the viscosity
(Eq.~\ref{eq:anisotropy_corrections0}) breaks down. (For example higher order
terms in Eq.~\ref{eq:anisotropy_corrections0} become important.  Additionally,
for $\kappa_{\LDA}\sim1$, $\mu/\omega_z\sim 2.7$ and shell effects, although
somewhat weak in the unitary Fermi gases~\cite{Forbes:2012yp}, may also become
important.) But by gradually increasing $\omega_z$ from $\omega_z \sim 2 \pi \times
10^4$rad/s to $\omega_z \sim 2 \pi \times 77000$ rad/s one could 
measure the tendency of $\eta_{xz}$ to decrease.

The damping rate for the scissor mode has been measured in the BEC-BCS
crossover region for weakly anisotropic traps in~\cite{PhysRevLett.99.150403}.
It will be interesting to see how the damping rate changes as $\omega_z$ is
increased.

On the other extreme, damping of the breathing and the radial quadrupole mode
(both insensitive to $\eta_{xz}$) was measured in the 2D Fermi
gas~\cite{PhysRevLett.108.070404}. It will be interesting to study the scissor
mode in these traps for smaller $\omega_z$.

  We hope our experimental colleagues in the cold atoms community will find our
proposal interesting and explore anisotropic viscosities in trapped
unitary fermions.

We especially thank M. Randeria for sharing his valuable insights. We also
acknowledge interactions with K. Damle, Shubhadeep Gupta, S. Minwalla, T.
Sch{\"a}fer, N. Trivedi, and D. G. Yakovlev. SPT acknowledges support from the
DST, Government of India.  We acknowledge support from the Infosys Endowment
and the DAE, Government of India. 

\bibliographystyle{ieeetr}
\bibliography{ucver1}

\end{document}